\documentclass[aip,jcp,reprint]{revtex4-1}
\usepackage{amsmath}
\usepackage{graphicx}
\usepackage{subcaption}

\begin{document}

\title{Translocation of short and long polymers through an interacting pore}
\author{F Piguet}
\author{D P Foster}
\email{damien.foster@u-cergy.fr}
\affiliation{Laboratoire de Physique Th\'eorique et Mod\'elisation (CNRS UMR 8089), Universit\'e de Cergy-Pontoise, 2 ave A. Chauvin 95302 Cergy-Pontoise cedex}


\begin{abstract}
    We perform two-dimensional Langevin dynamics simulations of  electric-field driven polymer translocation through an attractive nanopore. 
We investigate the effect of the location of the attractive region using different pore patterns.
This is found to have an impact on both the translocation time as a function of the chain length and on the polymer entry frequency. 
We qualitatively compare  our results to available experimental data.
\end{abstract}

\maketitle

\section{Introduction}
The passage of a molecule through a pore connecting two regions, known as translocation, is involved in many biological processes. 
Examples include the transport of RNA molecules through nuclear pores between the nucleus and the cytoplasm of the cell \cite{JMembrBiol-146} and the insertion of proteins in the membrane of cell components \cite{Cell-65}.
Today it is also the basis of technological applications, such as using pores as sensors for fast molecule sequencing \cite{PNAS-93,NatNanotech-4,PNAS-104} and as molecular sieves \cite{LabChip-8}. 

In 1996, Kasianowicz \textit{et al.} \cite{PNAS-93} established what is today the standard experimental method for studying the translocation of a macromolecule through a nanopore. 
A system composed of two distinct volumes connected only by the pore is filled with an ionic solution. 
The macromolecules to be translocated (polymers in our case) are placed in one or both volumes. 
Applying an electrical voltage across the pore causes a stable ionic current to flow 
through it. 
The presence of a polymer inside the pore reduces the available volume for the flow of ions, causing the measured current to decrease. 
Measuring the time intervals between current blockades allows the determination of the frequency at which a polymer enters the pore. The duration of the blockade gives 
the time it spends inside.

Through the numerous studies devoted to polymer translocation, a variety of parameters have been shown to have a possible influence on the process depending on the conditions:
If the pore diameter is smaller than the size of the polymer coil, the confinement of the polymer in the pore induces a loss of entropy which resists to its entry but helps its escape \cite{PRL-77-Su,JChemPhys-111}.
An external force may be applied to drive the polymer through the pore.
In the case of a charged polymer, the force results from the applied electric field\cite{PRL-85,PRL-86,Nanoscale-2}.
Such a driving force may also come from a difference of polymer concentration \cite{PRL-108} or solvent quality\cite{JChemPhys-126,JChemPhys-132} between the two compartments connected by the pore.
Depending on the charge of the pore walls, there might exist an electro-osmotic flow  which can either help or resist the polymer translocation. This effect is expected to be an important effect in synthetic pores\cite{PRE-78-Luan,JPhysCondMat-22}, which may be highly charged, but the relevance to biological pores, such as $\alpha$-hemolysin, is still debated\cite{JChemPhys-133,PNAS-107,PNAS-100}. 
Since the electro-osmotic flow depends on the external-field strength, it is difficult to isolate the effect of electro-osmosis 
when translocating charged polymers in an external field.
Finally the polymer may interact with the pore.
This interaction contains at least a steric part which prevents the polymer from penetrating the pore walls\cite{JChemPhys-119}.
In some cases specific repulsive or attractive interactions also exist between the polymer and the pore, which have been shown to have a great influence on the translocation process \cite{PRE-78-Luo,JChemPhys-130,PRL-99,JChemPhys-119}. 

Here we use a simple two-dimensional model for the translocation of a charged polymer moving through an interacting channel. In this study we shall concentrate on the effect of an attractive interaction with the pore and the effect of the 
external field.
The simplicity and the two-dimensional nature of our model 
prevents us from any direct comparison with experiments,  our results do however exhibit common qualitative features with experimental data, which will be discussed below. 


Translocation experiments use either biological or synthetic solid-state nanopores. 
The most commonly used biological pore is $\alpha$-hemolysin \cite{Science-274}. 
This pore exhibits both a geometrical and electrical asymmetry.
 It consists of two pieces of roughly equal length ($\approx$ 5~nm); a spacious vestibule connected to a narrower stem. 
In standard pH conditions the stem has a negatively charged ring at its outer extremity, 
while the rest of the pore is globally neutral \cite{JChemPhys-133}. 
Henrickson \textit{et al.} \cite{PRL-85} and Gibrat \textit{et al.} \cite{JPhysChemB-112} measured the entry frequency of negatively charged polymers (single-stranded DNA \cite{PRL-85} and dextran sulfate sodium \cite{JPhysChemB-112}) during translocation through $\alpha$-hemolysin as a function of the applied voltage. 
They considered both the case where polymers enter the pore from the vestibule side and the case where they enter from the stem side. 
The blockade frequencies were found to be always greater on the vestibule side in the voltage range they explored.
Two reasons were suggested to explain this result:
Firstly, the larger diameter of the vestibule entry reduces the loss of entropy necessary for the chain to enter the pore \cite{PRL-85,JPhysChemB-112}. 
Secondly, the negative extremity of the stem acts as a repulsive region for a negatively charged chain
 (while it would be an attractive region for a positive chain) \cite{PRL-85}. 
These two mechanisms contribute to make entry of negatively charged polymers easier from the  vestibule side of $\alpha$-hemolysin.
While Henrickson \textit{et al.} \cite{PRL-85} did not distinguish between these two contributions, Gibrat \textit{et al.} \cite{JPhysChemB-112} considered the easier entropic confinement to be mainly responsible for the greater polymer entry frequency on the vestibule side. 
Nevertheless, it has been shown experimentally that a modification of the charge distribution in the $\alpha$-hemolysin pore dramatically affects the entry frequency of ssDNA \cite{PNAS-105,wolfe2007,mohammad2008}. 
Such an effect has been observed regardless of which side the  polymers enter,  vestibule or stem side. 
The entry frequency was affected even when the charge was modified in a region of the pore that was not 
expected to influence chain insertion. 
The Debye length represents the distance over which an electrostatic interaction is thought to be screened by ions in solution\cite{Oukhaled:Epl:2008}. 
Surprisingly it was  found that a  modification the charge located further than a Debye length from the pore entry still had a significant effect on the polymer entry frequency. 
It has been suggested that the head part of the polymer could explore the pore more quickly and deeply than previously thought before being sufficiently inserted to provoke a detectable current blockade\cite{PNAS-105}. 
In this paper we test the entry of a polymer into a pore with a symmetrical shape but an asymmetrical interacting pattern. 
We show that an asymmetrical interaction alone can lead to significant differences 
in the 
probability of chain entry. This probability depends  
 on which side the polymer enters.

We also investigate the dependence of the polymer translocation time $\tau$ on the chain length $N$. 
Meller \textit{et al.} \cite{PRL-86} performed the translocation of single-stranded DNA through $\alpha$-hemolysin from the vestibule side. 
They identified two regimes in the chain length dependence of the translocation time, 
according to whether the polymer is sufficiently short to be accommodated as a whole inside the pore (stem) or not. 
In particular, they found that the mean translocation speed per monomer is constant for chains longer than the pore size. 
This observation was supported by a theoretical model based on the transport of the polymer through the free-energy landscape it experiences during the process \cite{JChemPhys-118-Sl} and by Langevin dynamics simulation \cite{PRL-96}. 
The model used contains the contributions of entropy and external driving force, as well as a polymer-pore interaction. 
The difference between the two regimes is attributed to an additional entropic contribution in the case of long chains from the portion of the chain that cannot be accommodated in the pore,  \cite{PRL-96}.

Reiner \textit{et al.} \cite{PNAS-107} reported the translocation of poly(ethylene glycols) (PEG) polymers through $\alpha$-hemolysin from the stem side. 
Whilst PEG is a neutral polymer, 
under certain circumstances it can
coordinate cations in ionic solution and therefore behave like a positively charged chain \cite{BiophysJ-95,CompTheorPolymSci-9,PNAS-107,Ghani2012}.
Reiner \textit{et al.} \cite{PNAS-107} considered these captured cations to be responsible for an attractive interaction between the PEG and the $\alpha$-hemolysin pore. 
In contrast to Meller \textit{et al.} \cite{PRL-86}, Reiner \textit{et al.} \cite{PNAS-107} described their results with a single free-energy barrier model, regardless of the chain length, short or long.
They assumed the translocation time to be overwhelmingly dominated by the time it takes for the polymer to escape from the pore. 
The barrier the chain must overcome to escape results from the dominance of the pore attraction over the external driving force and the entropy gain due to deconfinement. 
Both the single- \cite{PNAS-107} and two-regime \cite{PRL-86} models mentioned above take the pore as fully interacting, ignoring the asymmetry of the pore. 
However it has been shown experimentally \cite{PNAS-105,wolfe2007} and numerically \cite{PRX-2} that the location of the interacting region can also have a great effect on the translocation time. 
In this article we perform simulations in order to test the existence of a single regime or several regimes in the relation between the translocation time and the chain length. 
In particular we explore the effect of the position of the interacting region within the pore. 
We confirm that, whatever the pore pattern, a crossover exists in the $N$-dependence of $\tau$. 
We reexamine the experimental data from ref.~\cite{PNAS-107} and identify the crossover. 
As already noted in ref.~\cite{PRL-86}, the transition distinguishes between short chains, where all monomers can reside together in the pore, and longer ones. 
Nevertheless, we do not use the entropic argument previously used by Matysiak \textit{et al.} \cite{PRL-96} to explain the difference between the two regimes and propose our own interpretation. 

The chronology of a translocation event gives us some further insight into the mechanisms at work. 
In particular, in addition to the contributions of entropy, external force and pore attraction, we highlight the importance of a crowding effect in the polymer chain. 
We show that the collective movement of the monomers, which must move together in a one direction for the chain to progress, plays a significant role when the external driving force acting on the chain is not too strong. 

\section{Simulation method}

The results presented here come from two-dimensional simulations performed using the ESPResSO simulation package \cite{CompPhysCom-174}. 
The simulation system and the parameters values  follow closely the work of Luo \textit{et al.}\cite{PRE-78-Luo} and 
Cohen \textit{et al.} \cite{PRX-2}.

The polymer is modelled by a self-avoiding bead-spring chain, each bead representing a monomer. 
The excluded volume interaction between monomers is set by a purely repulsive truncated Lennard-Jones potential:
\begin{equation}\nonumber
  U_{mm}^{LJ}\left(r\right)=
  \begin{cases}
    4\epsilon_{mm}\left[\left(\frac{\sigma}{r}\right)^{12}-\left(\frac{\sigma}{r}\right)^{6}\right]  & \text{if }  r \le r_{mm}^{c}\\\\
    U_{mm}^{LJ}\left(r_{mm}^{c}\right) & \text{if }  r > r_{mm}^{c}\\
  \end{cases}
\end{equation}
where $r$ is the distance between consecutive monomers, $\epsilon_{mm}$ is the potential depth and $\sigma$ the monomer diameter. 
The cutoff distance $r_{mm}^c$ is set to the potential minimum to eliminate the attractive part, giving $r_{mm}^{c}=2^{\frac{1}{6}}\sigma$. 
The chain connectivity is maintained by a FENE potential:
\begin{equation*}
  U^{FENE}\left(r\right)= -\frac{1}{2}kr_{max}^2\ln\left[1-\left(\frac{r}{r_{max}}\right)^2\right]
\end{equation*}
where $k=30 \epsilon/\sigma^2$ is the spring stiffness and $r_{max}=2\sigma$ the maximum distance 
between consecutive monomers. 
These parameter values lead to a mean bond length equal to $\sigma$. All the beads are identical and carry unit charge.

\begin{figure}
  \includegraphics[width=0.9\linewidth]{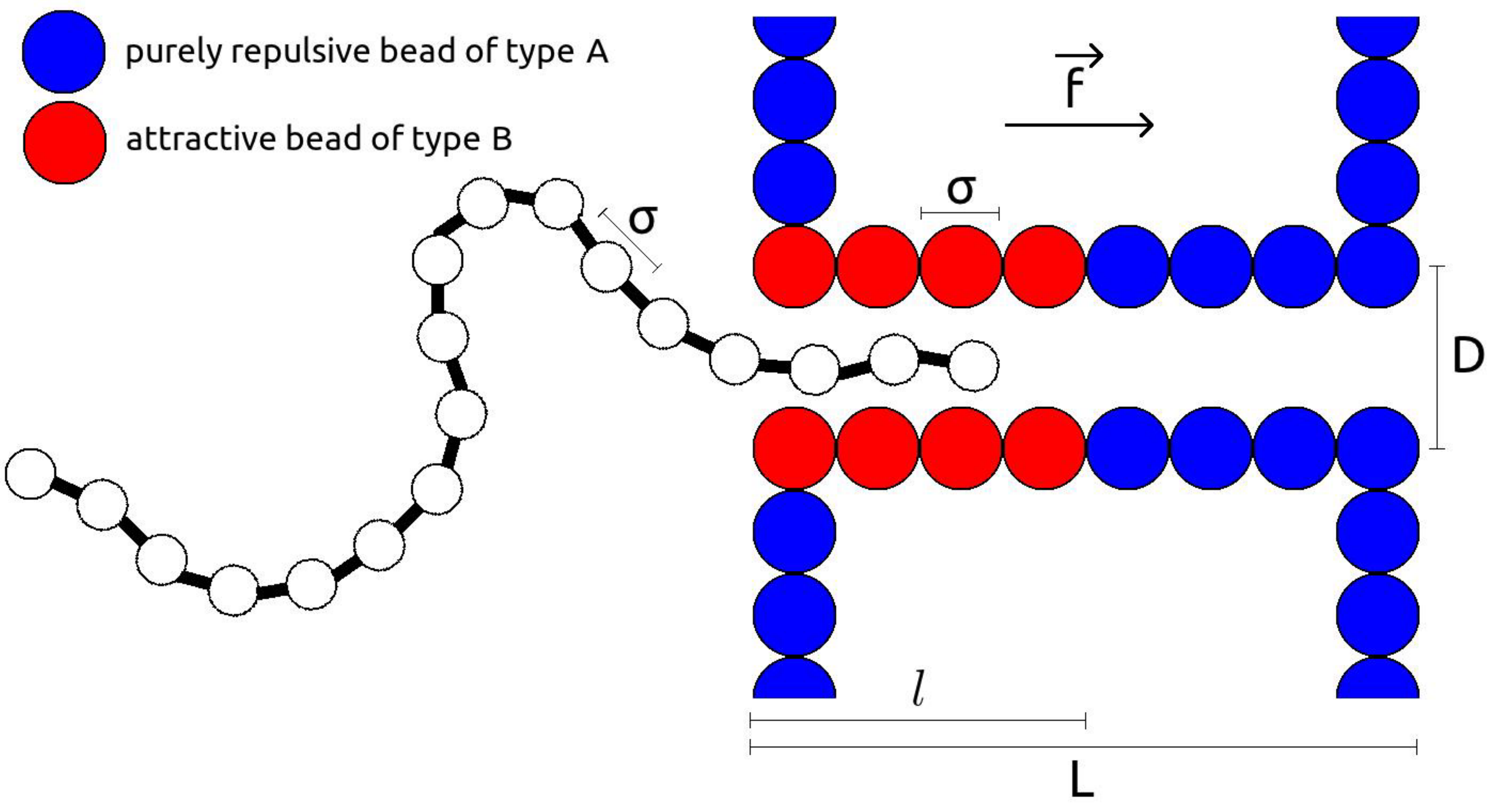}
  \caption{A view of the simulation system in the case of a polymer entering on the attractive side of the pore. In this example the pore length is reduced to $8 \sigma$ for sake of simplicity.}
  \label{sim_system}
\end{figure}

\begin{figure}
  \includegraphics[width=0.9\linewidth]{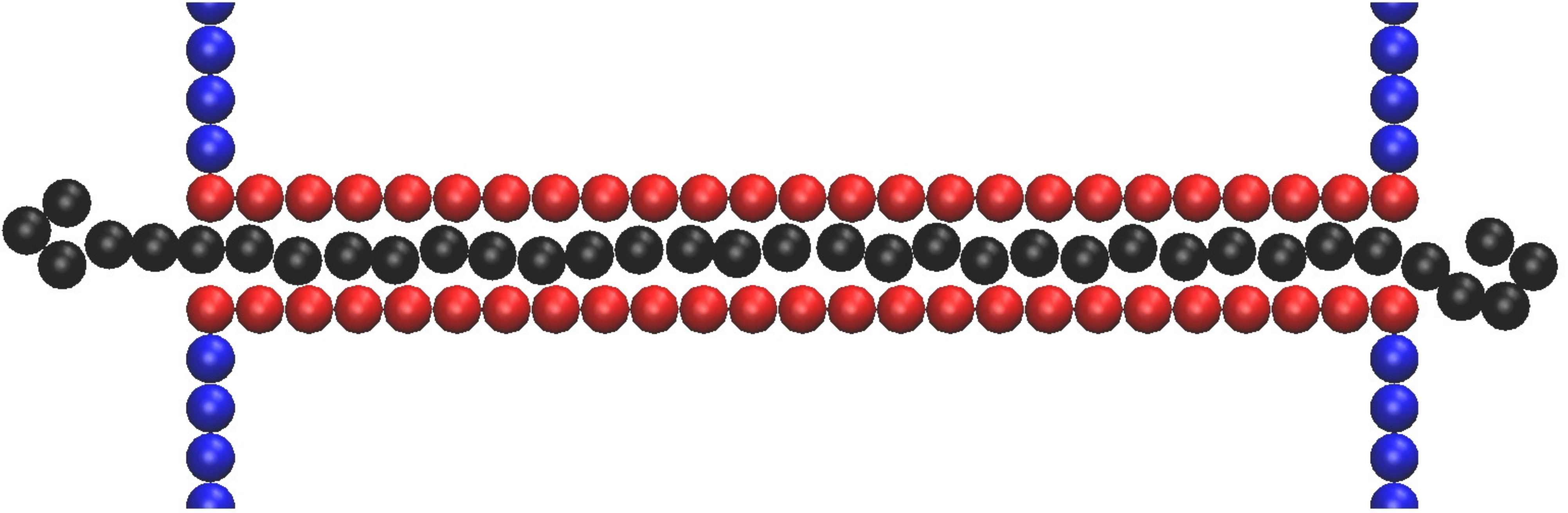}
  \caption{A snapshot from the ESPResSO simulation, showing the single file  transport of the polymer  through the pore.}
  \label{indian_file}
\end{figure}

The pore and membrane walls consist of static beads separated 
from each other 
by a distance $\sigma$. 
The pore length is $L=25\sigma$. 
The distance  $D=2.25\sigma \approx 2 \times 2^{\frac{1}{6}}\sigma$ between the centres of the wall beads ensures that there  is a strong energetic penalty for any significant deviation of a monomer from the central axis of the pore, ensuring 
single-file translocation (see figure~\ref{indian_file} for snapshot). This restriction is not a problem, since the model may be considered to be a coarse grained version of the real polymer, and as long as the pore is narrow enough, the model monomers may be considered to be blobs in the de Gennes picture. 
We define two types of bead (A and B) which have different interactions with the polymer. 
All the beads in the membrane walls are of type A. 
The channel walls are symmetric 
about a median axis
and can be made of beads of type A or B (Fig. \ref{sim_system}).

The interaction between a monomer and a bead of type A or B is again described by a truncated LJ potential but with parameters ($\epsilon_{mA}$,$r_{mA}^c$) and ($\epsilon_{mB}$,$r_{mB}^c$) respectively. 
A type A bead acts as a purely repulsive site for a monomer; we set $\epsilon_{mA}=\epsilon_{mm}$ and $r_{mA}^c=r_{mm}^{c}$. 
A type B bead is an attractive site, thus we choose $\epsilon_{mB}=\epsilon_{mm}$ and $r_{mB}^c=2.5\sigma$ to ensure the existence of an attractive part. 
Finally, each monomer inside the pore experiences a driving force $\vec{f}$ directed along the pore axis simulating the effect of the external electric field.

The equation of motion of a monomer $i$ can be written:
\begin{equation*}
  m\ddot{\vec{r}}_i = -\vec{\nabla}U_i + \vec{f} - \zeta \dot{\vec{r}}_i + \vec{\eta}
\end{equation*}
where $m$ is the monomer mass, $U_i$ is the sum of the interactions from the other monomers of the chain and from the wall beads on the monomer $i$, $\vec\eta$ is a random force mimicking the thermal agitation caused by collisions with the solvent molecules, and related to the friction coefficient $\zeta$ by the fluctuation-dissipation theorem \cite{PR-83}. 
For $\vec\eta$, we use the Langevin thermostat provided by the simulation package to set the system temperature $T$. 

Every run begins with an equilibration phase, during which one end of the polymer is fixed at the pore entrance while the rest of the chain is free to fluctuate. 
As we do not consider any hydrodynamic interactions, we take the time for the polymer to relax to equilibrium as the Rouse time $\tau_R \sim N^{1+2\nu}\ $\cite{JChemPhys-21}. 
The first bead is then released and the process is monitored. 
Each data point reported here results from averaging over 2000 independent simulations, at least in the region of interest. 

In the molecular dynamics simulations all quantities are expressed in terms of an energy  scale $\epsilon_{mm}$,  
a length scale, $\sigma$, and a mass scale, $m$. In terms of these parameters, the time scale is defined by the time unit
$t=\left(m\sigma^2/\epsilon_{mm}\right)^{1/2}$. The temperature is given by $k_BT=0.85\epsilon_{mm}$ and the frictional coefficient  $\zeta$ is set to $0.7 m/t\ $. The driving force on a monomer in the pore is set to 
$f=1.0\epsilon_{mm}/\sigma$. The parameter values were chosen to be in the range typically used in the literature\cite{PRE-78-Luo,PRX-2}. It is difficult to relate the results directly to an experimental setup of a polymer with a particular voltage 
and charge per monomer, since the simulation groups different types of interaction together. For example an electro-osmotic flow within the pore would, at the level of our simulation, also appear like a charged polymer in a field. The interest of the simulations here are to gain some qualitative insight. However, if one where to take  $\epsilon_{mm} =$ 35 meV \cite{JPhysChemB-113}, $\sigma =$ 0.35 nm \cite{BiophysJ-84} and $m =$ 44 amu as the PEG monomer binding energy,
length and mass respectively.  We would have a temperature of about 20$^\circ$C and a force of about 13 pN per monomer in  the pore.

\section{Results and discussion}

\subsection{Probability of Entry}

The frequency of blockade events  depends on both the electric field and on which side the polymer enters from\cite{PRL-85,JPhysChemB-112}.
The time between blockades may be broken down into three parts. 
The first is related to the time it takes a polymer to diffuse in the solution to the region of the pore entrance. 
For narrow pores, such as considered here, once the polymer has arrived near the pore, it must position itself such that one end is presented to the pore entrance before it can enter. The second part corresponds to the time for this to happen.
Lastly, for there to be a blockade, the polymer must overcome any free-energy barrier and enter  the pore for a time 
long-enough for there to be a measurable drop in the current. This blockade event may or may not correspond to an actual translocation.
The first step should not depend on the pore or the applied electric field. The second step could depend on the electric field strength, as the field is expected to extend weakly beyond the entrance of the pore\cite{2012arXiv1206.1967F}, and is sensitive to hydrodynamic effects\cite{JChemPhys-120,JChemPhys-123}.
However, one expects that only the third step depends on the detailed nature of the interior of the pore. The third
step is the most amenable to simulation, and the first two steps are both independent of which side the pore enters. In what follows we concentrate on this third step.

In order to test the effect of an asymmetric interaction on the polymer entry frequency, each wall of the pore we use consists of an attractive part made of type B beads, on a length $l=13$, completed by $L-l=12$ purely repulsive type A
beads. 
In the rest of this paper, when using this asymmetric pore, the pore side made of type B beads 
will be naturally referred to as the pore attractive side, while its other end  will be referred 
to as the repulsive side, for the sake of simplicity. 



\begin{figure}
\begin{subfigure}[b]{0.9\linewidth}
	\includegraphics[width=\textwidth,clip]{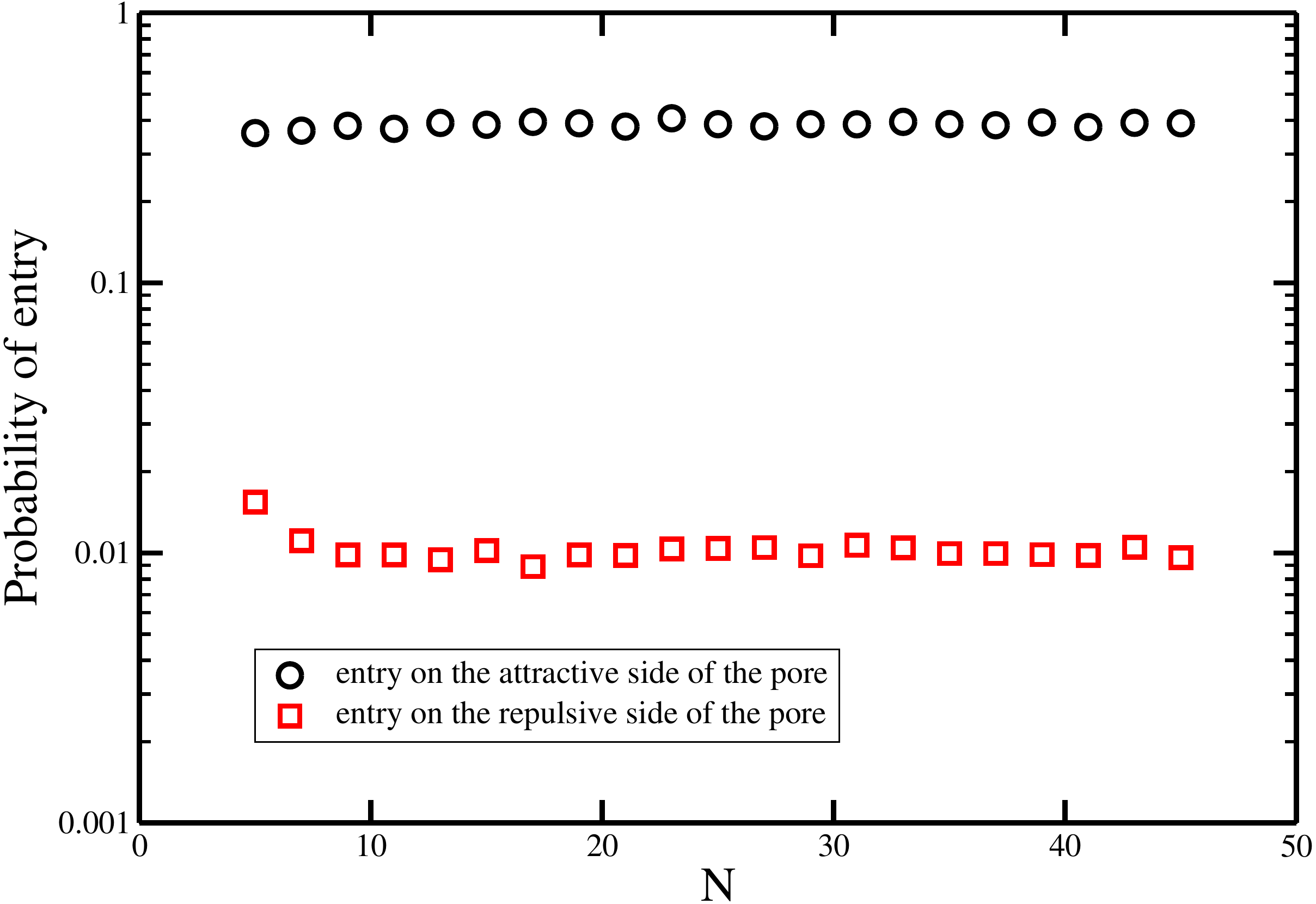}
\caption{Entry probability}\label{proba_entry}
\end{subfigure}	

\begin{subfigure}[b]{0.9\linewidth}
  \includegraphics[width=\textwidth,clip]{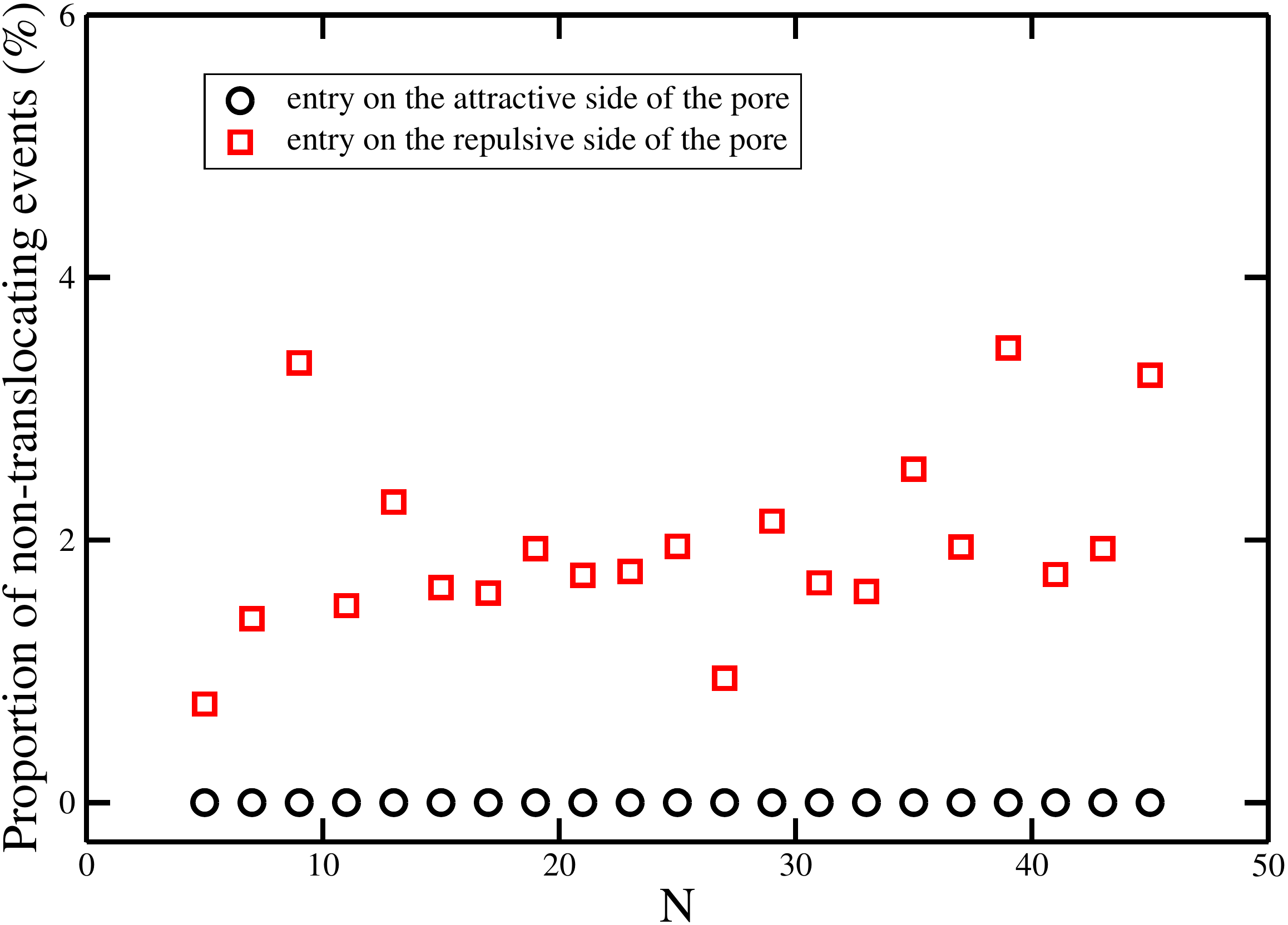}
  \caption{Proportion of polymers which fail to translocate}\label{prop_cb}
  \end{subfigure}
  \caption{The curves for both the probability and proportion of non-translocating events are shown for both sides of entry as a function of the polymer length}
\label{probas}
\end{figure}

The probability that the polymer enters the pore from an initial configuration with a monomer in the pore entrance has been extracted from our simulations for several chain lengths $N$ and one value of the external force $f$, in both the case where the polymer initially faces the pore attractive side and the case 
where the polymer faces the 
pore repulsive side (Fig. \ref{proba_entry}). 
It must be noted that in experiments a criterion is adopted to distinguish between current blockades due to the presence of a polymer in the pore and current variations due to the noise of the system. 
One or both of the following criteria are commonly used: the current drop must be sufficiently strong and/or its duration must be sufficiently long \cite{PNAS-107}. 
These criteria in themselves cannot distinguish between polymers which translocate and polymers which 
exit from the side they entered. 
In order to perform an analysis comparable to what is done experimentally, we shall only consider that the polymer succeeds in entering the pore when it dwells
more than a minimal time $t_{min}$
inside the pore.
We choose $t_{min}=80t$ to obtain a selection criterion 
comparable with experiments of Reiner {\it et al.} \cite{PNAS-107}.

Our results (Fig. \ref{proba_entry}) show that, for each chain length, the probability of observing a current blockade is greater when the polymer is initially located near the pore attractive entrance rather than near the repulsive one. 
It is natural to expect that an attraction from the pore entrance facilitates chain insertion. 
These results demonstrate that the location of the polymer-pore interaction, decoupled from any geometrical asymmetry, is sufficient to cause great variations in the polymer entry frequency. 
Furthermore, in the range of chain lengths we explore, the probability of entry is roughly independent on the chain length,
in agreement with results of Luo et al\cite{JChemPhys-134-Luo}. 
This reveals that the entropic cost to confine the first monomers and dwell more than $t_{min}$ in the pore does not depend on the chain length in this range. 
It is also of interest to check the proportion of failed events (where the polymer exits from the pore on the same side it entered) in the total number of recorded current blockades as a function of the polymer entry side (Fig. \ref{prop_cb}). 
It appears that, for the chosen value of $t_{min}$, all the events where the polymer entered on the pore attractive side are translocations. 
This is not the case on the repulsive entry side, 
although an overwhelming majority of events (more than 95\%) are still translocations. 
It is likely that the proportion of failed events is involved in the nonmonotonic behavior of the residence time that has been experimentally observed for very long chains, as has been suggested \cite{PRL-97}. 

\subsection{Translocation time}
The translocation process is divided into three steps: filling, transfer and escape. 
To allow an easier comparison between the different pore patterns we use, the different stages are delimited by the position of the polymer relative to the pore interacting region. 
The \textit{filling} stage ($\tau_1$) begins with the entry of the first monomer into the pore and ends either when the interacting zone has been filled (long chain) or when the whole chain has entered this region (short chain). 
Then the \textit{transfer} stage ($\tau_2$) lasts until the chain end enters the interacting part of the pore (long chain) or until the head monomer reaches the exit of this region (short chain). 
Finally, the \textit{escape} stage ($\tau_3$) corresponds to the pore emptying. 

\subsubsection{Uniformly attractive pore}
\begin{figure}
  \begin{subfigure}[b]{0.9\linewidth}
  \includegraphics[width=0.9\textwidth,clip]{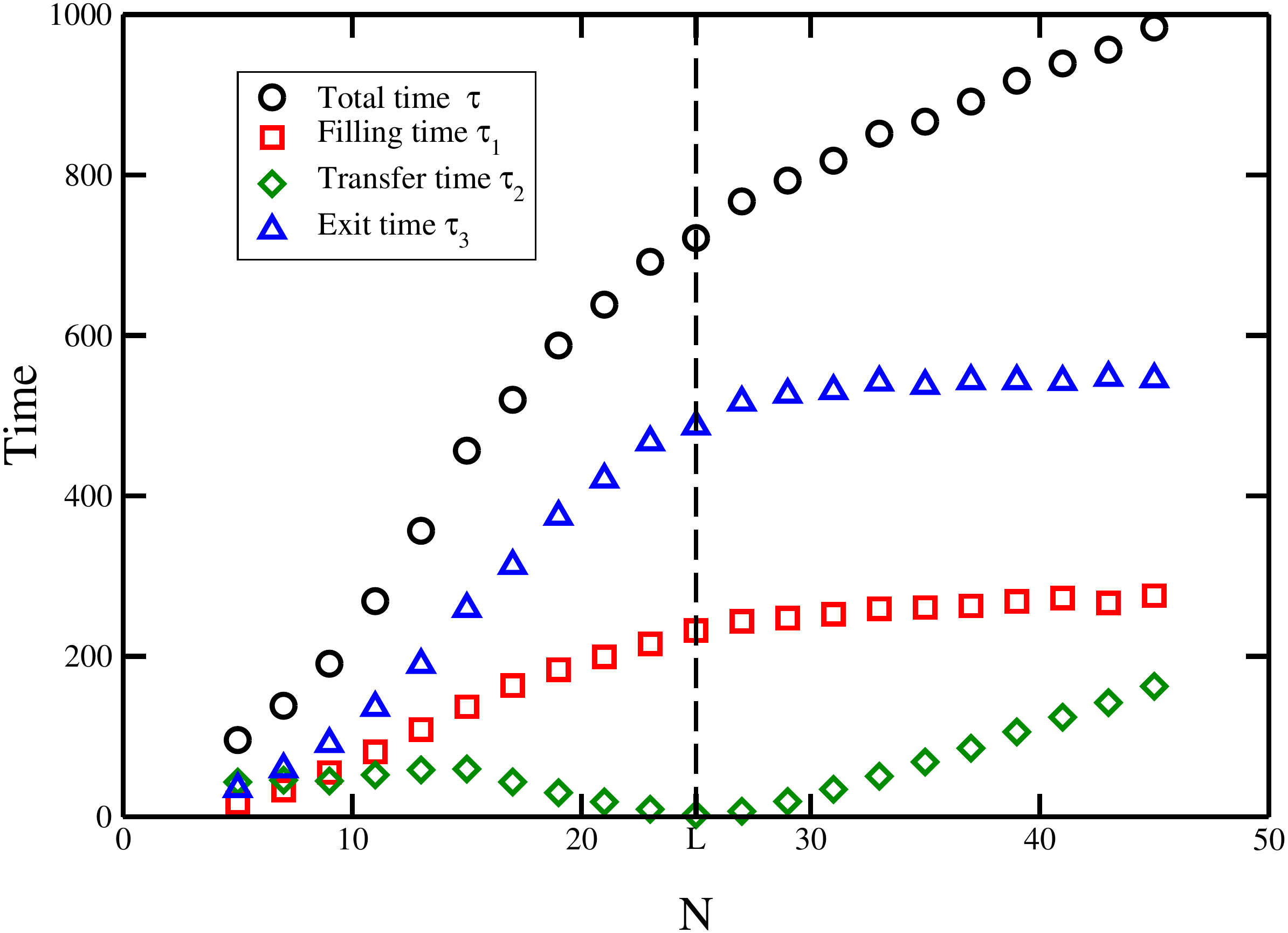}
  \caption{Uniformly attractive pore}\label{uniform}
  \end{subfigure}
  
  \begin{subfigure}[b]{0.9\linewidth}
	\includegraphics[width=0.9\textwidth,clip]{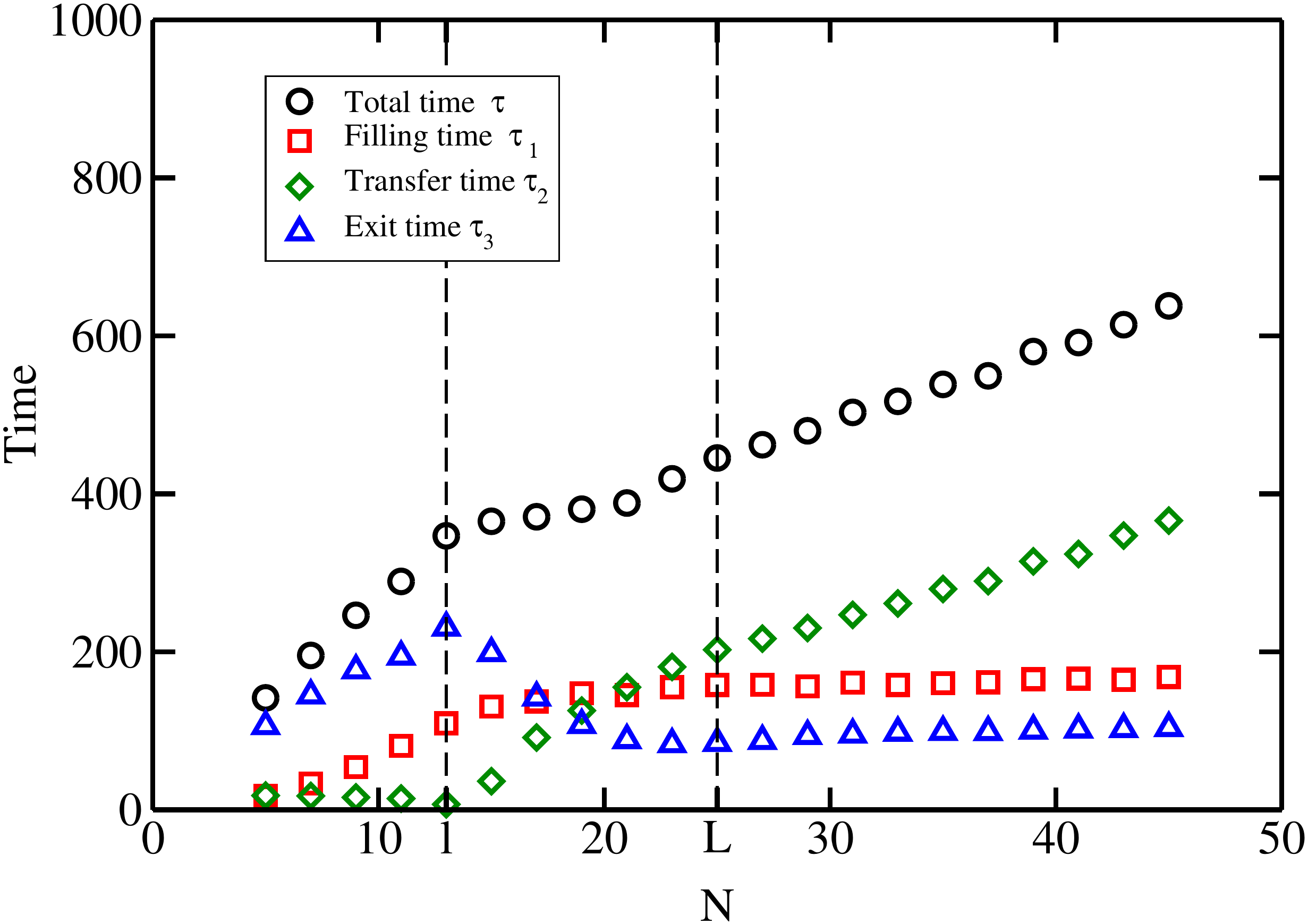}
	\caption{Pore with attractive section on side of polymer entry}\label{attractive_entry}
\end{subfigure}
  
  \begin{subfigure}[b]{0.9\linewidth}
  \includegraphics[width=0.9\textwidth,clip]{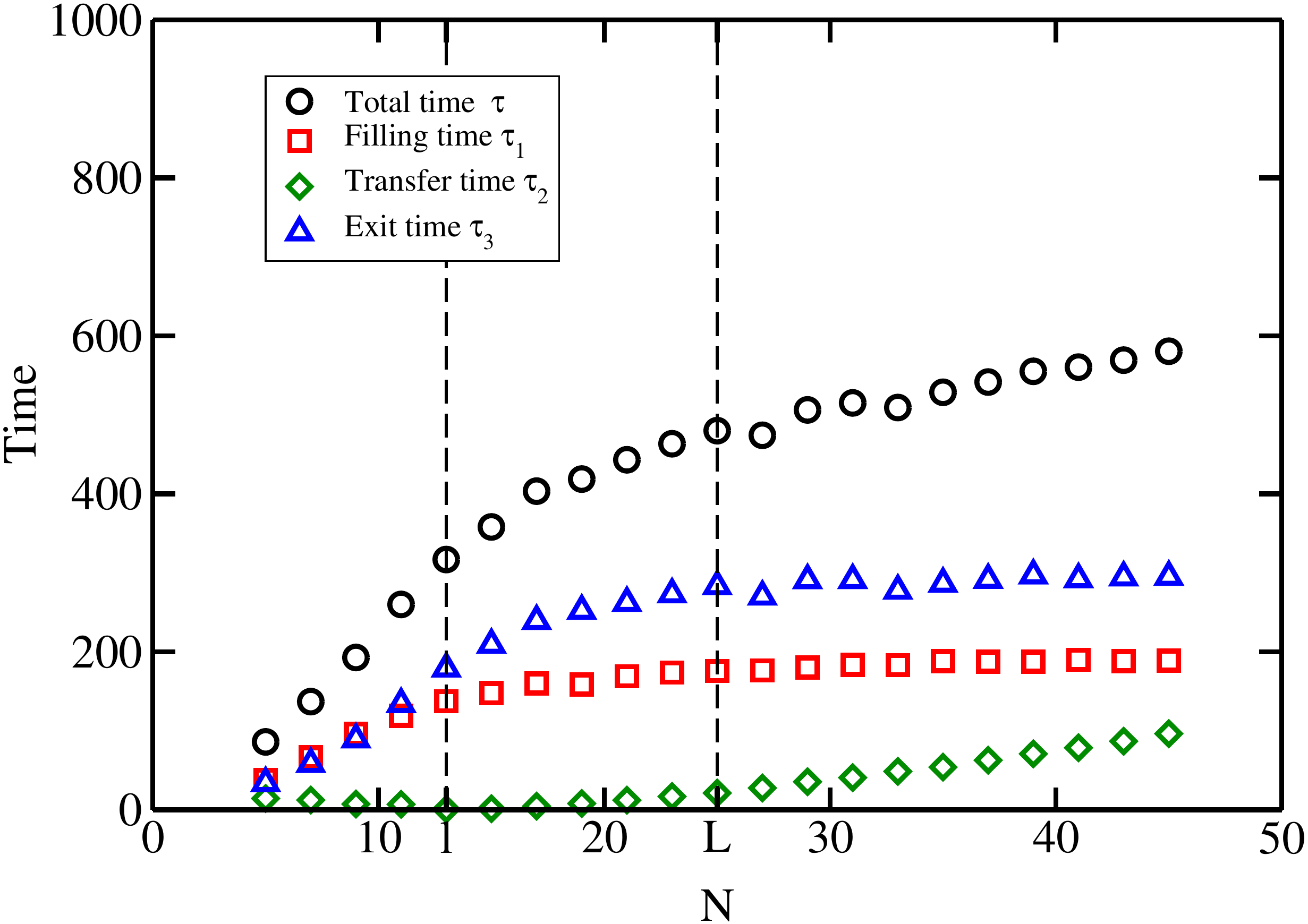}
  \caption{Pore with repulsive section on side of polymer entry}\label{repulsive_entry}
  \end{subfigure}
  \caption{Translocation time and its components versus chain length.}
\label{Results_sim}
\end{figure}

In order to highlight the main features of the translocation process, we first study the case of a pore whose walls are uniquely made of attractive beads. 
The mean translocation time $\tau$ and its components $\tau_{1,2,3}$ as a function of the chain length $N$ are shown in Fig. \ref{uniform}. 
Because in this case the pore (length $L$) and its interacting part (length $l$) are equal, we shall refer to them both as \textit{the pore}. 
Two regimes clearly appear for $\tau$ separated  by a transition region corresponding to the pore size $L$, with a linear behaviour for $N>L$. 
The $\tau_{1,2,3}$ curves reveal that the first regime is dominated by the escape stage $\tau_3$ (and filling stage $\tau_1$ to a lesser extent), while the second regime is dominated by the transfer step $\tau_2$.

In order to enter the pore, the chain must experience a force sufficiently strong to counterbalance the loss of entropy due to confinement. 
This force is due to the external driving force and the pore attraction. 
The filling time $\tau_1$ increases with $N$ until $N=L$ because of an increasing number of beads to confine. 
All chains longer than $L$ contain more monomers than the pore can accommodate. 
For all of these chains, the filling stage corresponds to the confinement of the same number of monomers, which is the maximum number of monomers  the pore can accommodate. 
Thus the filling time $\tau_1$ remains roughly constant for chains longer than the pore size. 
This reveals that, at least for long chains, the chain length does not play any significant role in the filling process which begins with one polymer end facing the pore entrance. 
In particular, it does not significantly affect the entropic cost to fill the pore. 

During the transfer stage, the entropic and pore attraction contributions remain constant. 
This is obvious when the chain is sufficiently short to be accommodated as a whole in the pore. 
In this case, the transfer stage is just the transport of the chain inside the pore, from one end to the other. 
This begins when the tail end of the chain leaves the pore entrance and finishes when its head end
reaches the front exit. 
An increase of the chain length reduces the distance it
needs to cover inside the pore and increases the external force 
it feels, causing the transfer time $\tau_2$ to decrease until $N=L$. 
For chains longer than the pore size, the transfer stage is the transport of the chain tail that has not been confined during the filling step. 
Constant entropic and pore attraction contributions result from the 
entry of a new
monomer at one end of the pore 
is coincides with the exit of a monomer at the other end.
This suggests that the entropic variation when confining/deconfining a monomer is 
roughly independent of the chain length outside the pore \cite{Macromolecules-41}.
This is compatible with our observations in the range we consider. 
Thus a constant driving force, due to the action of the external force on a constant number of confined monomers, is equilibrated by the attraction from the pore walls and friction with the solvent.
This leads to a constant speed of transfer, i.e. a linear increase of $\tau_2$ with $N$. 

\begin{figure}
  \includegraphics[width=0.9\linewidth,clip]{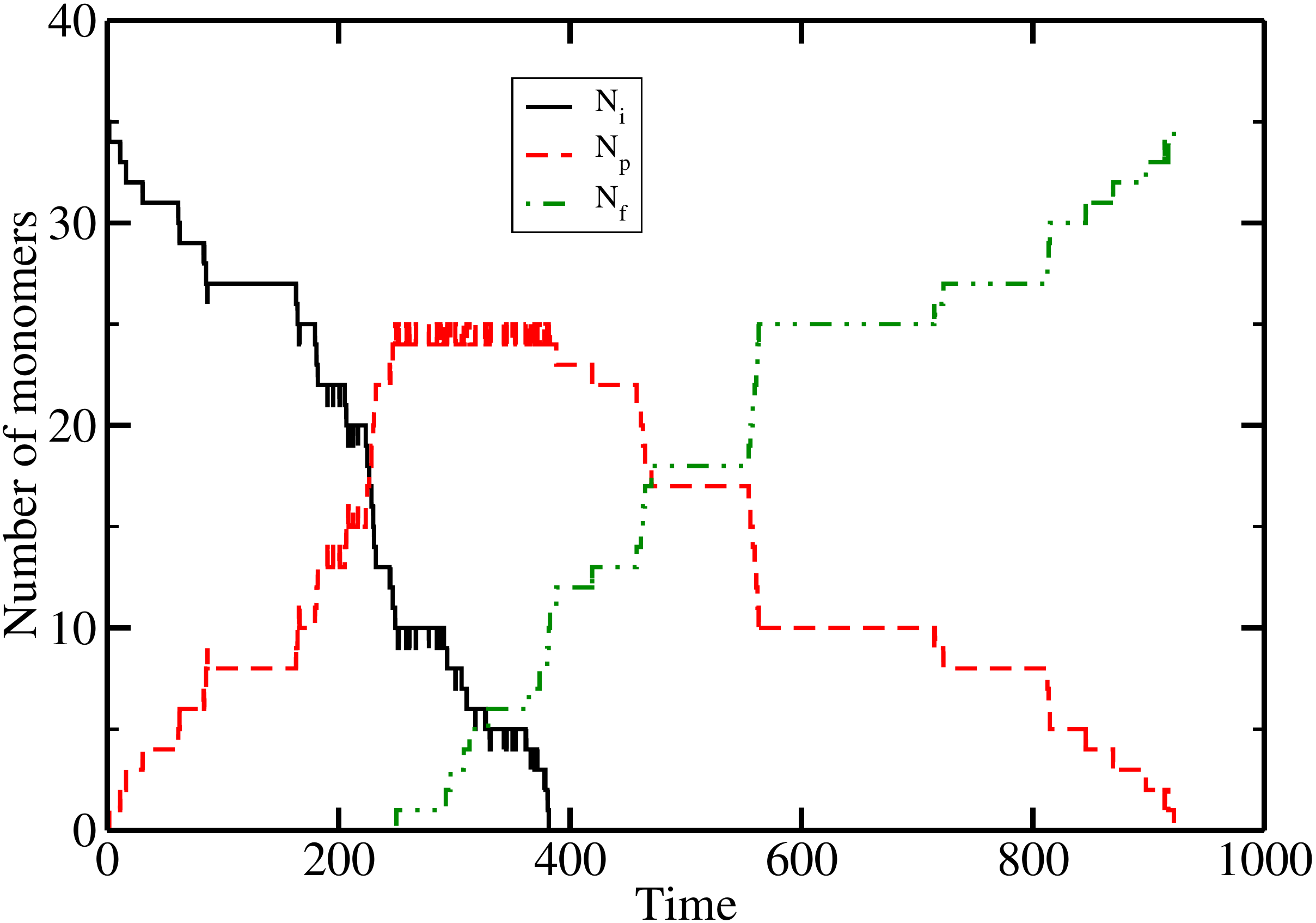}
  \caption{Time evolution of the chain's position during a translocation event: $N_i$ monomers have yet to enter the pore, $N_p$ are inside the pore and $N_f$ have already translocated. The pore is uniformly attractive and the polymer length is $N=35>L$.}
\label{traj}
\end{figure}

The emptying stage $\tau_3$ is similar to the filling one, except that now the gain of entropy due to deconfinement and the external driving force must overcome the pore attraction for the polymer to escape. 
As for the filling time, the escape time first increases as the number of monomers to release increases until $N=L$. 
Then, for longer chains, emptying a full pore requires roughly the same time, independently of the chain length. 
\\\\
Analysing the chronology of a translocation event gives us some further insight into the translocation process (Fig. \ref{traj}). 
With our parameter values, the external force is sufficiently strong compared to the pore attraction to avoid the existence of any significant energy barrier. 
Adding the entropic contribution to obtain the polymer free energy could 
reveal a free-energy barrier in the filling stage, as it resists polymer confinement, but not in the emptying stage where it works with the electric field in helping the polymer escape.
We find that considering an entropy cost of $k_B\ln\mu$ for the confinement of a monomer
(where $\ln \mu$ is of order one) is not sufficient to create  an entry barrier. 
The correction term for the polymer segment of $N_{out}$ monomers yet to enter is 
$\left(\gamma-1\right)\ln\left(N_{out}\right)$ with $\gamma=0.70$, chosen to take into account the presence of the 
membrane wall\cite{RevModPhys-65}. This correction would tend to make the entropy cost  even smaller. 
It should be noted that these expressions for the entropic contribution may not be appropriate for two reasons: firstly they are scaling relations, valid for long chains, and secondly they assume that equilibrium conditions apply.

Whilst we have just shown that, with our parameters, no  
barrier exists in the pore emptying stage, and that it seems unlikely to find one in the filling stage, the translocations we recorded are not continuous drifts. 
On the contrary, we observe that, whatever the stage of the process, it consists of  
a stepwise stick-slip type motion (see Fig. \ref{tram}).  A similar observation was made in all atom molecular dynamics simulations\cite{ACSNano-6,JChemPhys-127} 
During the filling stage, the number of beads 
entering the pore in a single forward movement increases with time (from one to eight monomers at once). 
The longer the part of the chain inside the pore, the greater the external force it feels, while the entropic cost and energy gain from pore attraction when a new monomer is confined remain roughly constant. 
Then, as the chain progresses through the pore, the stronger external driving force allows 
larger groups of monomers to enter in one go.
Despite this increasing driving force, the waiting time between two jumps to fill the pore first increases until the first eight monomers have entered, and then decreases.
While the number of beads inside the pore is small, the external force acting on it is weak
and the progress of the chain through the attractive channel 
is hindered by the fact that for one monomer to move forwards it must wait for the monomers ahead to move first 
(crowding effect). 
As the confined part of the chain increases, every new step requires a longer time since more monomers have to diffuse at the same time. 
Once a threshold number of monomers have been confined ($\approx 8$ in our example), the driving force acting on the polymer overcomes the crowding effect. The monomers ahead move fast enough to free space for those behind,  
allowing larger groups of beads to enter in shorter time intervals. 
This interpretation correlates with the observed inflexion of $\tau_1$ in the $N<L$ regime: as the pore attraction dominates the progression of short chains ($N \lesssim 15$, Fig. \ref{uniform}), confining every new bead takes an increasing time until the external force becomes the dominant effect, after which the mean entry time per new monomer decreases.

The same story, but in reverse order, describes the escape stage $\tau_3$. 
Now the driving force and the gain of entropy due to deconfinement act against the pore attraction. 
If the confined part of the chain is sufficiently long, the beginning of the emptying process is dominated by the driving force. 
After a threshold number of monomers have been released from the pore, the confined part of the chain is sufficiently short for the crowding effect to be dominant. 

%
\subsubsection{Influence of the location of the attractive region}
We perform  simulations in order to investigate the effect of an asymmetric interaction between the polymer and the pore on the translocation time. 
We consider the half-interacting pore we previously used, with an attractive region of length $l=13$. 
We simulate both the case where the polymer initially faces the pore attractive side 
and the case where it faces 
the repulsive side (Fig. \ref{attractive_entry} and \ref{repulsive_entry}). 

Clearly the mean translocation time is shorter, since the attractive region is shorter. The interesting effect is the difference in behaviour depending on which end the polymer enters.
When the polymer enters the attractive side, a new regime appears, 
for chain lengths comprised between the length of the interacting region ($l$) and pore size ($L$). 
With our parameter values, the translocation time stays nearly constant in this range. 
On both sides of this region, no qualitative change can be observed compared to the case of a uniform interaction. 

Considering the $\tau_i$ components in the case where the polymer enters on the attractive side of the pore (Fig. \ref{attractive_entry}), we first note that the filling time $\tau_1$ now reaches a roughly constant value for chains longer than the length of the interacting part of the pore. 
This results from the definition of $\tau_1$, which corresponds to the filling of the pore interacting region. 
More spectacular changes concern the transfer and escape times. 
Between $l$ and $L$, the curve of the transfer time $\tau_2$ exhibits a new non-linear part, while the escape
time $\tau_3$ decreases. 
These two results have the same origin. 
As the interacting region now covers only a part of the pore, a part of the chain can be located in a pore region where it feels the external force but not the pore attraction. 
When the polymer enters the pore on its attractive side, such a situation does not occur during the filling stage. 
During the transfer stage, the head portion of the chain that has already escaped from the attractive zone but not from the pore experiences an increasing driving force as it moves forward. 
This makes the confinement of every new bead easier and easier, leading to a decreasing mean additional transfer time per monomer for $l<N<L$. 
Similarly, the emptying stage for chains with $l<N<L$ is facilitated by the action of the external force on the head part of the chain, already outside the interacting zone but still inside the pore. 
This additional help, compared to chains with $N<l$, causes $\tau_3$ to decrease. 
Globally, with our parameter values, the help provided by an extra bead interacting with the external force and not with the pore is such that the reduction of the escape time balances the additional transfer time. 
This leads to a nearly constant value of the translocation time for chain lengths between $l$ and $L$. 
For chains longer than the pore length, once the first monomer has escaped from the pore, the number of monomers experiencing the external force remains constant.
While the pore is full, the entropic and pore interaction contributions also remain constant. 
The transfer stage is thus completed by the linear transport of the end part of the chain which has not yet been confined and $\tau_2$ increases linearly for $N>L$. 
Finally, because all the chains with $N>L$ must extract the same number of monomers from the pore in order to empty it, we observe once again a constant value for the escape time.

The results obtained when the polymer enters on the repulsive side of the half-interacting pore are shown in 
Fig. \ref{repulsive_entry}. 
Even if a region still exists where the external force acts on the monomers
unbalanced by the pore attraction,
 all the curves exhibit a monotonic behaviour, in contrast with the previous case.
 The resulting extra force now contributes to the $\tau_{1,2}$ steps.
 The acceleration it provides when one monomer is added to the chain is not sufficient to balance the additional filling and transfer times endured. 

If the total chain length did not have any significant impact on the chain progression
once the chain is long enough,
we would have expected $\tau_3$ to be constant for $N>l$. 
This is not the case: $\tau_3$ still increases above $l$, 
albeit more and more slowly reaching a constant asymptote.
We conclude that the total chain length does have an influence here. 
We examine the converse situation, i.e. the entry stage on the pore attractive side (Fig. \ref{attractive_entry}). 
Even if the effect is smaller, 
$\tau_1$ also shows a slightly increasing part above $l$. 
We deduce that the effect of the chain length is not entropic in nature, because it slows both the polymer entry and escape. 
It is rather a manifestation of the crowding effect between the monomers, which dominates the chain progression at the beginning of the filling stage and at the end of the emptying stage. 
Our observation reveals that this crowding effect not only involves
 the confined monomers but also the monomers located near the pore entrance, whose movements are correlated with those of the inserted part of the chain. 
Monomers located further in the chain structure outside the pore, if they exist, are sufficiently decorrelated from those inside the pore to have a negligible influence on their movements. 
Hence the crowding effect saturates above a certain chain length. 

\subsubsection{Comparison with experiment}

We must first recall that experimental data in references~\cite{PRL-86,PNAS-107} 
relate to the time the polymer spends inside the pore, and do not distinguish on which side the polymer escapes.
 The discussion above only dealt with
 translocation events. 
In this section we show our results
for the dwell time as a function of chain length for
all events where at least a portion of the chain resides more than a minimum time $t_{min}$ inside the pore.
Once again we chose $t_{min}=80t$ so as to be consistent with 
the experimental criterion.
The results for all events were quantitatively and qualitatively the same as before, reflecting the fact that we are in the strong field limit, and virtually all the polymers which enter the pore translocate.
Thus in what follows we use the results shown on Fig. \ref{Results_sim} to compare with experimental work. 

\begin{figure}
  \includegraphics[width=0.9\linewidth,clip]{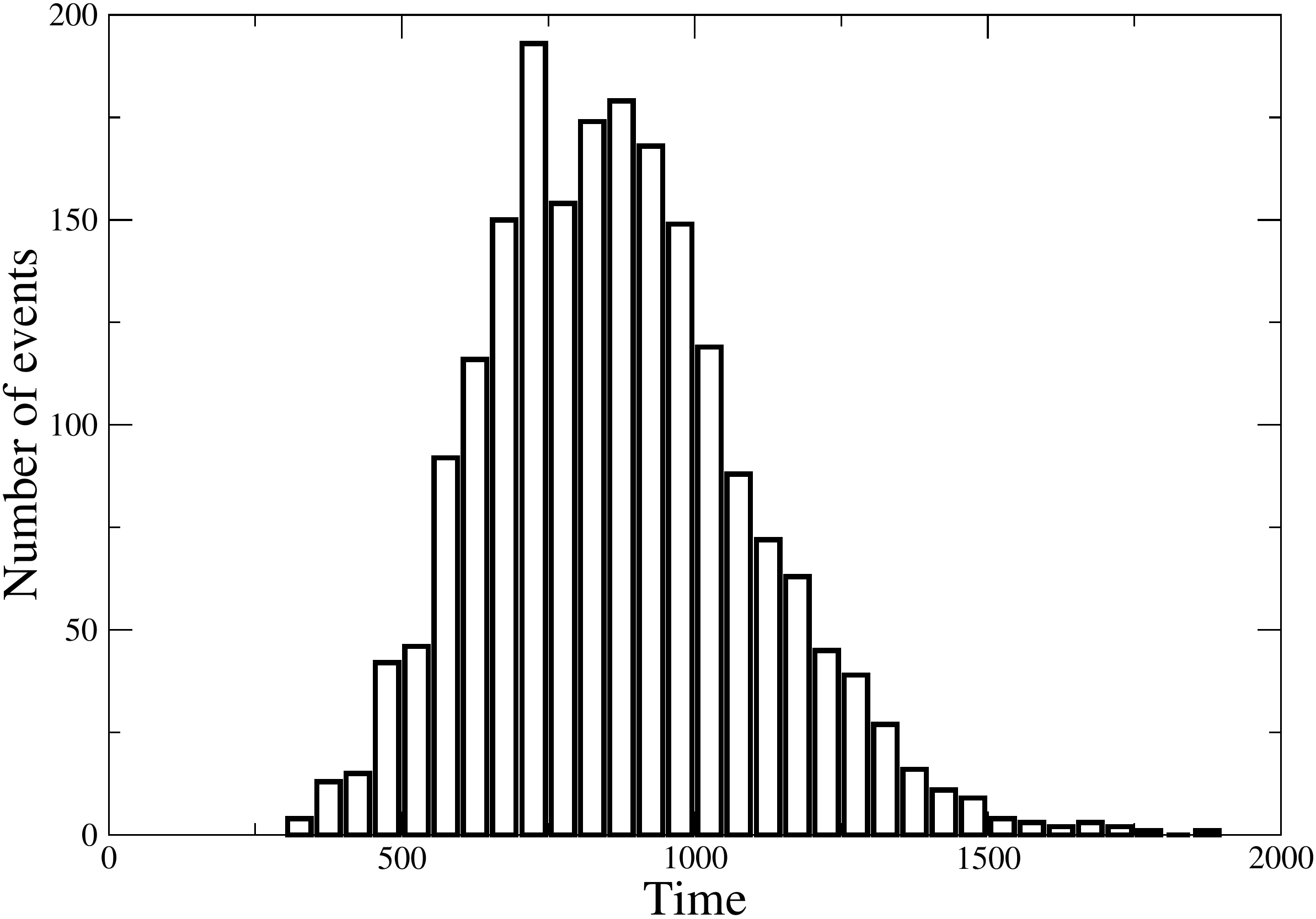}
  \caption{A typical distribution of the translocation times from our simulations. The pore is uniformly attractive and the polymer length is $N=35>L$.}
  \label{Distrib}
\end{figure}

Our results on the translocation of a charged polymer through an interacting pore are in good qualitative agreement with the experimental observations of Meller \textit{et al.} \cite{PRL-86} on the transport of single-stranded DNA (4 to 100 bases) through $\alpha$-hemolysin. 
The distribution of translocation times from our simulations (Fig. \ref{Distrib}) is qualitatively similar with the experimental distribution reported in ref~\cite{PRL-86}.
This indicates that our choice of parameters is compatible with the conditions of the experiments we refer to.
Meller \textit{et al.} \cite{PRL-86} identify a crossover in the chain-length dependence of the translocation time for a strand length $N^* \approx 12$ DNA bases. 
The DNA bases translocate in single file, just as in our simulations. 
They also observe a crossover for the mean pore current as a function of the chain length at the same value of $N^*$. 
For longer chains, the translocation time 
increases 
linearly 
with the chain length, and the current attenuation remains constant at its minimal value. 
As the current through the pore decreases when the number of monomers inside the pore increases, this indicates that the crossover corresponds to the maximal chain length which can be accommodated 
 in the pore and contributes to the current drop. 
But a value of 12 DNA bases corresponds to a linear chain length of 4.8 nm \cite{PRL-86}, i.e. only half of the $\alpha$-hemolysin pore length. 
This suggests 
that the current drop is mainly due to the presence of a polymer in a limited region of the $\alpha$-hemolysin pore. 
The duration of the current blockades corresponds to the transport of the polymer through this particular region. 
This region can accommodate a maximum of 12 DNA bases, and thus has a maximum length of roughly 5 nm considering the confined part of the chain as totally unfolded. 
While the two parts of the $\alpha$-hemolysin pore (vestibule and stem) have roughly equal lengths of 5 nm, the narrowness of the stem part causes greater current drops in the presence of a polymer inside. 
Moreover molecular dynamics simulations  have shown that the essential of the potential drop occurs through the stem part of $\alpha$-hemolysin\cite{BiophysJ-88}.
Thus we can reduce the translocation through $\alpha$-hemolysin to the crossing of the stem channel, which is assimilated to the pore we used in simulations. 
Our simulation results confirm the experimental observation that a crossover occurs for the behaviour of the translocation time with the chain length when the chain becomes too long to be accommodated in the pore.


\begin{figure}
  \includegraphics[width=0.9\linewidth,clip]{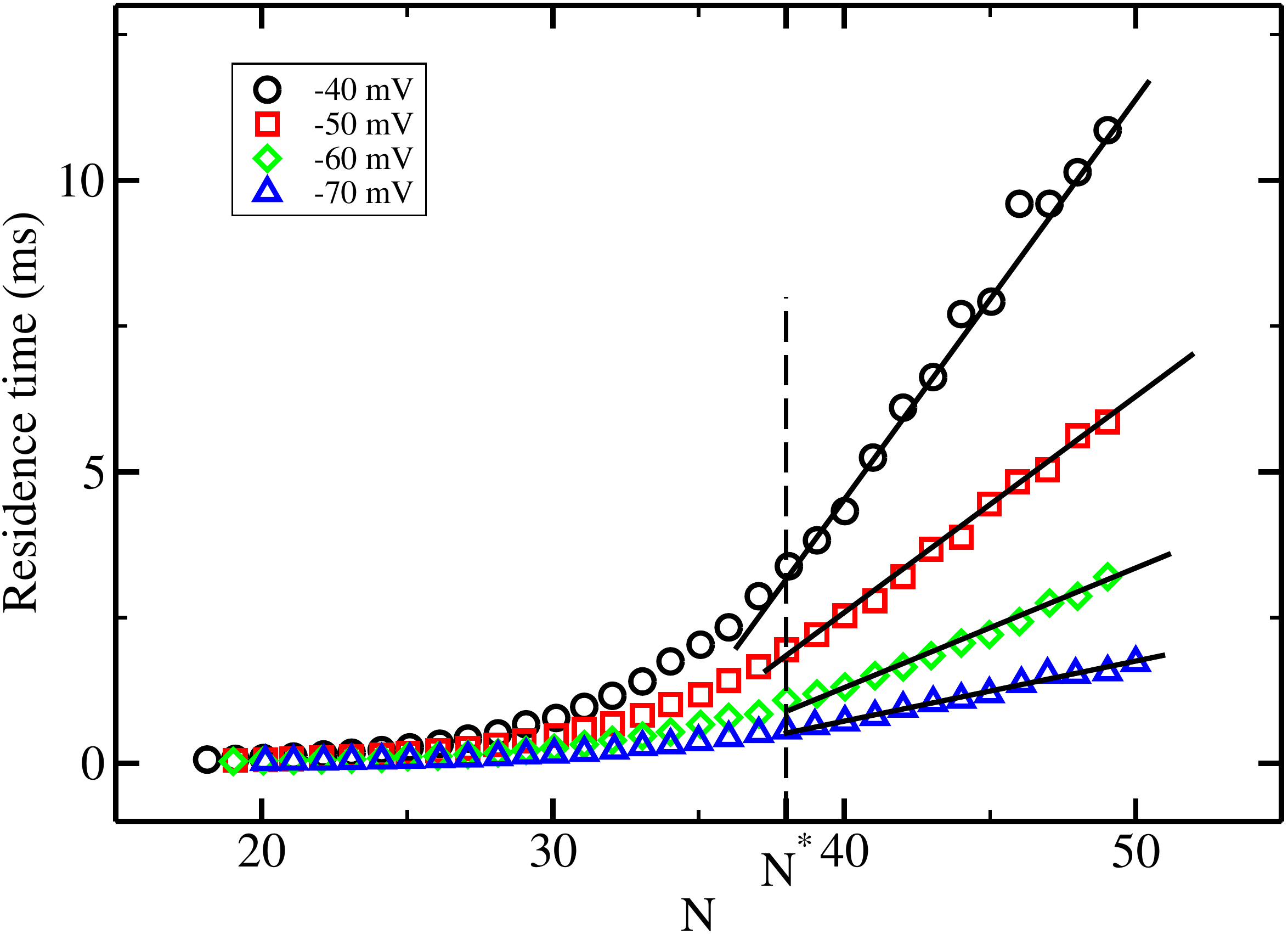}
  \caption{Experimental results on PEG polymer residence time in $\alpha$-hemolysin as a function of the polymer length taken from Figure 5 D in Reiner {\it et al.}\cite{PNAS-107}.  The curves become linear beyond $N^*$, the solid straight lines are place to help guide the eye.}
  \label{Results_exp}
\end{figure}

We also compare our conclusions to the experimental results of 
Reiner \textit{et al.} \cite{PNAS-107} on the translocation of PEG (19 to 50 monomers) through $\alpha$-hemolysin (Fig. \ref{Results_exp}). 
These results were interpreted with a single-regime model,
i.e. without a crossover in the polymer length\cite{PNAS-107}. 
Reexamining their data in the light of our results, we identify the crossover between a non-linear and a linear behaviour of the translocation time with the chain length for each voltage value (Fig. \ref{Results_exp}). 
This transition occurs for a polymer of length $N^* \approx 38$ PEG monomers. 
By contrast with Meller \textit{et al.} \cite{PRL-86}, the mean pore current as a function of the chain length still decreases for chains longer than this threshold value \cite{PNAS-107}. 
Taking a PEG monomer length as $\sigma \approx$ 0.35 nm \cite{BiophysJ-84}, the linear length of a chain of 38 PEG monomers is approximately equal to 13 nm. 
Translocation through $\alpha$-hemolysin is essentially 
equivalent to the polymer transport through the stem channel. Unfolded, such a chain would greatly exceed the length of the stem  (5~nm).  
The diameter of the smallest part of the $\alpha$-hemolysin ($\approx$ 1.5 nm~\cite{BiophysJ-88}) is about four PEG monomers lengths, we expect the confined polymer to be coiled as a linear chain of ``blobs'' \cite{PGdG-Scaling} 
under the voltages used in ref.\cite{PNAS-107}. 
The size of a blob is determined by the channel diameter (vestibule or stem depending of its position in the pore).
For chains made of more than 38 PEG monomers, the additional monomers 
in the vestibule 
add a significant
additional  contribution to the drop in current.
This contrasts with the case of Meller \textit{et al.} \cite{PRL-86}, where the DNA was also extended in the vestibule. 
This might explain the continued decrease of the current 
for chains longer than $N^*$ observed by
Reiner \textit{et al.} \cite{PNAS-107}.


\section{Conclusion}
Through the simulations, we have shown the importance of the location of a polymer-pore interaction in the electric-field-driven translocation of a charged polymer. 
When only one of the two ends of the pore is interacting with the polymer, the probability for the chain to enter the pore is naturally greatly affected when the pore side on which the polymer tries to enter is reversed. 
This is in agreement with experimental work testing the effect of a modification of the charge distribution inside the $\alpha$-hemolysin pore on the polymer entry frequency from a given pore side \cite{PNAS-105,wolfe2007,mohammad2008}. 
As well as the geometrical asymmetry of the $\alpha$-hemolysin pore, the asymmetry of the interaction may play a significant role in the differences observed experimentally in the polymer entry frequency 
depending on which end the polymer enters\cite{PRL-85,JPhysChemB-112}.
The position of the interaction also causes quantitative as well as qualitative changes in the dependence of the polymer translocation time $\tau$ on the chain length $N$. 
The qualitative behaviour of $\tau$ with $N$ may serve as a criterion to resolve the location of the polymer-pore interaction in experiments, although the effects may be small compared to the thermal noise.
We have demonstrated that a crossover exists in the relation between $\tau$ and $N$, regardless of the interacting pattern. The crossover corresponds to the maximum chain length that can be accommodated in the pore. 
This crossover has been identified in experimental results for single-stranded DNA \cite{PRL-86} and for PEG \cite{PNAS-107} 
translocation through $\alpha$-hemolysin. 
We have shown that the transition from a non-linear to a linear dependence of the translocation time on the chain length does not arise from the entropy
  as suggested by Matysiak \textit{et al.} \cite{PRL-96}. 
It is rather the consequence of
 constant velocity transport of a portion of a long chain. 
The question of very long chains was out of the scope of this paper, at least for reason of excessive computational time.
With our choice of parameters, we expect the linear behaviour of the translocation time with the chain length to be still valid for very long chains as it has been observed in experiments \cite{PRL-86,Electroph-23} and numerical studies \cite{PRE-78-Luo,PRL-96,JChemPhys-119} under appropriate conditions.
Some of the qualitative factors of the dwell time were attributed to a crowding effect between monomers, which hinders the chain's progress through the pore when the external force acting on it is not too strong, as is the case in many experiments. It would be interesting to study this point further.

\acknowledgements
The authors would like to thank J. Pelta and A. Oukhaled for many useful discussions and introducing us to the problem of polymer translocation. 

%

\end{document}